\documentclass[journal,twoside,web]{ieeecolor}
\usepackage{generic}
\usepackage{url}
\usepackage[colorlinks,linkcolor=blue]{hyperref}
\usepackage{cite}
\usepackage{amsmath,amssymb,amsfonts}
\usepackage{algorithmic}
\usepackage{graphicx}
\usepackage{textcomp}
\usepackage{booktabs}
\usepackage{color}
\newcommand{\tr}{\textcolor[rgb]{1,0,0}}

\def\BibTeX{{\rm B\kern-.05em{\sc i\kern-.025em b}\kern-.08em
    T\kern-.1667em\lower.7ex\hbox{E}\kern-.125emX}}
\begin{document}
\title{RFormer: Transformer-based Generative \\ Adversarial Network for Real Fundus Image\\ Restoration on A New Clinical Benchmark}
\author{Zhuo~Deng, Yuanhao~Cai, Lu~Chen, Zheng~Gong, Qiqi~Bao, Xue~Yao, Dong~Fang, \\ Wenming~Yang, \IEEEmembership{Senior Member, IEEE}, Shaochong~Zhang, Lan~Ma
\thanks{This work was supported in part by the Shenzhen Bay Laboratory and the Shenzhen International Science and Technology Information Center. }
\thanks{Zhuo~Deng, Yuanhao~Cai, Zheng~Gong, Qiqi~Bao, Wenming~Yang and Lan~Ma are with Shenzhen International Graduate School, Tsinghua University, Shenzhen 518055, P.R. China (email: \{dz20, cyh20, gz20, bqq19\}@mails.tsinghua.edu.cn, \{yang.wenming, malan\}@sz.tsinghua.edu.cn).}
\thanks{Lu~Chen, Shaochong~Zhang, Dong~Fang, Xue~Yao are with Shenzhen Eye Hospital affiliated to Jinan University, Shenzhen 518040, P.R. China (email: \{ chenludoc, shaochongzhang\}@outlook.com, dora.eye@hotmail.com, 18925257121@163.com).}
\thanks{Co-first authors: \textit{Zhuo~Deng}  and  \textit{Yuanhao~Cai}.}
\thanks{Corresponding authors: \textit{Lan~Ma}, \textit{Shaochong~Zhang}, and \textit{Wenming~Yang}.}}
\maketitle

\begin{abstract}
Ophthalmologists have used fundus images to screen and diagnose eye diseases. However, different equipments and ophthalmologists pose large variations to the quality of fundus images. Low-quality (LQ) degraded fundus images easily lead to uncertainty in clinical screening and generally increase the risk of misdiagnosis. Thus, real fundus image restoration is worth studying. Unfortunately, real clinical benchmark has not been explored for this task so far. In this paper, we investigate the real clinical fundus image restoration problem. Firstly, We establish a clinical dataset, Real Fundus (RF), including 120 low- and high-quality (HQ) image pairs. Then we propose a novel Transformer-based Generative Adversarial Network (RFormer) to restore the real degradation of clinical fundus images. The key component in our network is the Window-based Self-Attention Block (WSAB) which captures non-local self-similarity and long-range dependencies. To produce more visually pleasant results, a Transformer-based discriminator is introduced. Extensive experiments on our clinical benchmark show that the proposed RFormer significantly outperforms the state-of-the-art (SOTA) methods. In addition, experiments of downstream tasks such as vessel segmentation and optic disc/cup detection demonstrate that our proposed RFormer benefits clinical fundus image analysis and applications. The dataset, code, and models will be made publicly available at \url{https://github.com/dengzhuo-AI/Real-Fundus}.
\end{abstract}


\begin{figure}[!t]
\centerline{\includegraphics[width=1\columnwidth]{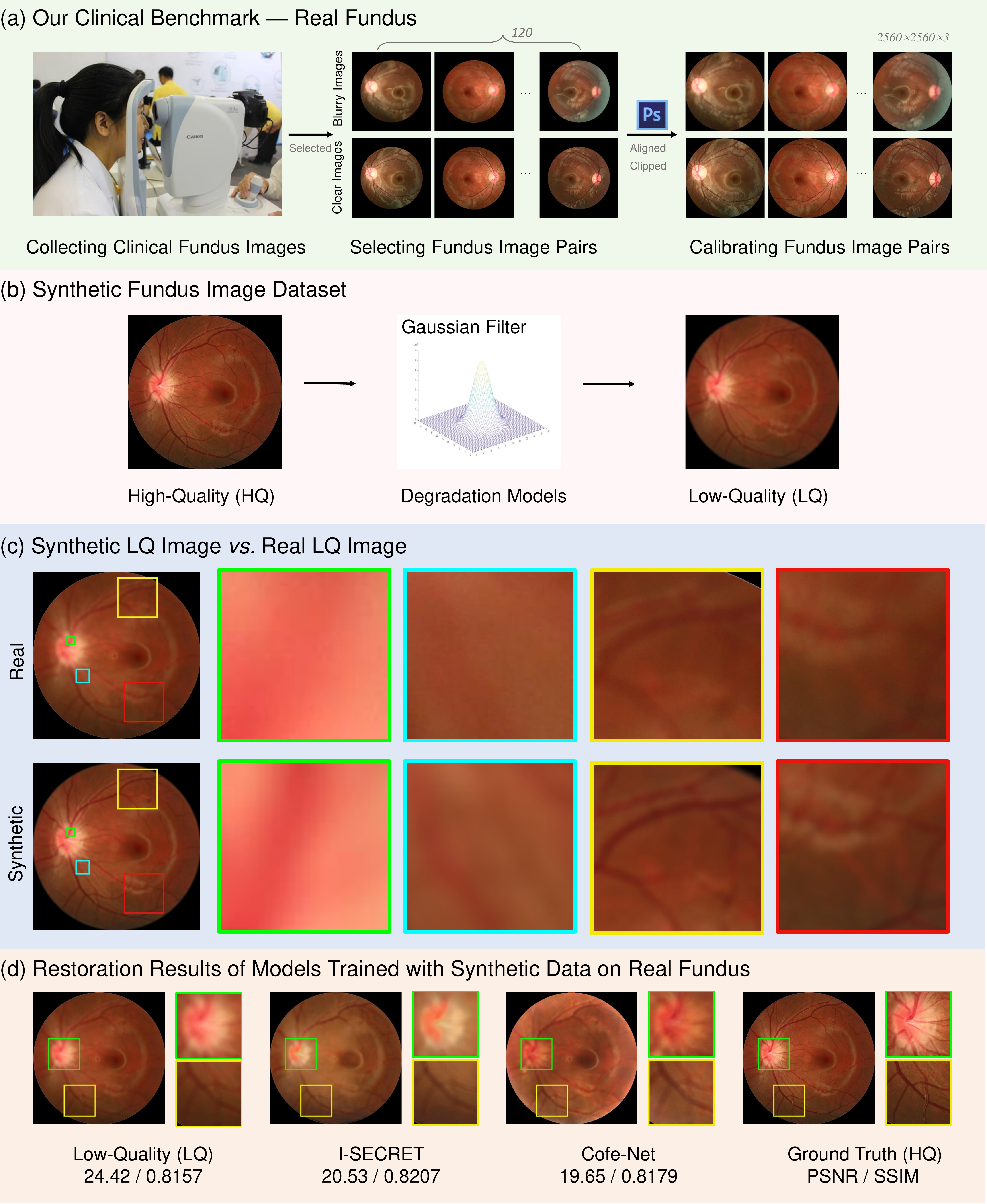}}
\caption{ \textbf{Our Real Fundus \emph{vs.} Synthetic Dataset.} (a) The pipeline of establishing our clinical fundus image benchmark, Real Fundus (RF). (b) Artificial degradation models are used to synthesize low-quality (LQ) fundus images from their high-quality (HQ) counterpart. (c) Comparisons of synthetic LQ fundus image and our real clinical LQ image. (d) Restoration results of models trained with synthetic data on RF. The two CNN-based methods, I-SECRET~\cite{cheng2021secret} and Cofe-Net~\cite{shen2020modeling}, fail to reconstruct the real clinical degraded fundus images.}
\label{dataset}
\end{figure}

\section{introduction}
\label{sec:introduction}
\IEEEPARstart{D}{ue} to the safety and cost-effectiveness in acquiring, fundus images are widely used by ophthalmologists for early eye disease detection and diagnosis, including glaucoma~\cite{chen2015glaucoma,mojab2019deep,liao2019clinical}, diabetic retinopathy~\cite{majumder2021multitasking,he2020cabnet,hua2020convolutional}, cataract~\cite{dong2017classification,zhang2019automatic}, and age-related macular degeneration~\cite{peng2019deepseenet,burlina2016detection}. However, different equipments and ophthalmologists pose large variations to the quality of fundus images. A screening study of 5,575 patients found that about $12 \%$ of fundus images are of inadequate quality to be readable by ophthalmologists~\cite{philip2005impact}. We analyze the factors causing the degradation in real fundus image capturing. \textbf{Firstly}, patients, especially infant patients, do not cooperate with the capturing process of fundus images. Specifically, most patients are reluctant to undergo pupil dilation, which causes poorly lit and blurred fundus images. Besides, infant patients usually can not resist the eye-closing reflex caused by a bright light during flash photography. \textbf{Secondly}, in practice, spatial pixel misalignment, color, and brightness mismatch are inevitable due to the changes in light conditions and misoperations of inexpert ophthalmologists. 
\textbf{Thirdly}, high-quality (HQ) fundus images can be collected in hospitals of developed areas using high-precision fundus cameras. However, these equipments are expensive and unaffordable for hospitals in some remote areas of under-developed or developing countries. As a result,  low-precision and portable fundus cameras are used to capture low-quality (LQ) fundus images. 
These LQ fundus images easily mislead the clinical diagnosis and lead to unsatisfactory results of downstream tasks like blood vessels segmentation. Various biomarkers of the retina ($e.g.$, hemorrhage, microaneurysm, exudate, optic nerve and optic cup) are essential in different diseases. Therefore, it is necessary to ensure the prominence and visibility of each marker for precise clinical diagnosis. Thus, when LQ  fundus images are captured in clinical diagnosis, ophthalmologists often repeat dozens of shots until HQ fundus images are obtained. Nonetheless, this repeated capturing process harms patients, degrades hospital efficiency, prevents reliable diagnosis of ophthalmologists, and impacts automated image analysis systems.   

\begin{figure}[!t]
\centerline{\includegraphics[width=\columnwidth]{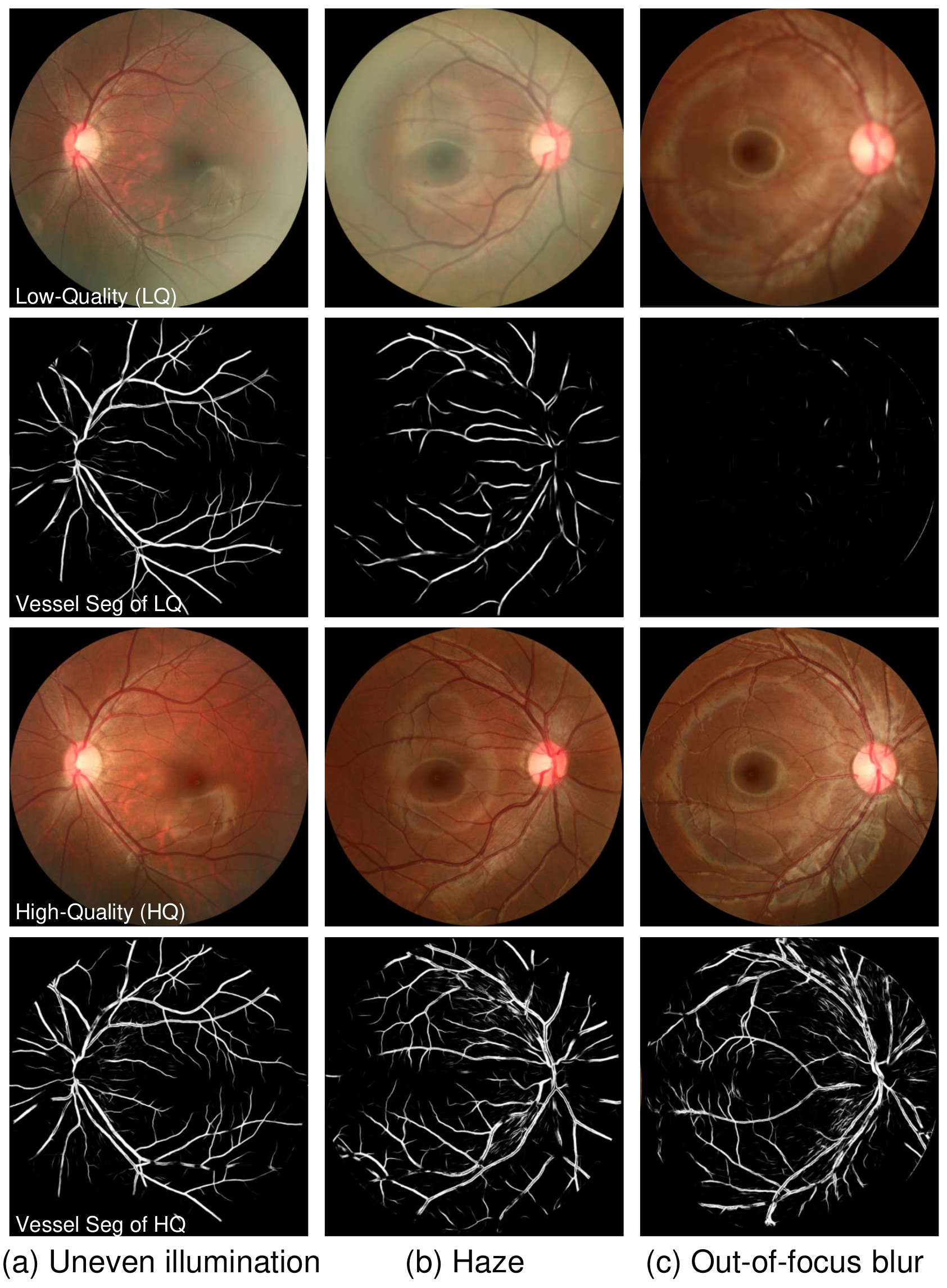}}
\caption{ \textbf{Fundus images of different degeneration types and their blood vessel segmentation results.}
(a) Uneven illumination. (b) Haze. (c) Out-of-focus blur. From top to bottom are the low-quality (LQ) fundus images, blood vessel segmentation of the LQ fundus images, the ground-truth high-quality (HQ) fundus images, and blood vessel segmentation of the ground-truth HQ fundus images.}
\label{dd}
\end{figure}

We observe the clinical fundus images and find that the main degradation types of LQ images  include out-of-focus blur, motion blur, artifact, over-exposure, and over-darkness. Compared with other types of degradation, blur, especially out-of-focus blur, poses the most severe threat to image analysis and clinical diagnosis. An example is shown in Fig.~\ref{dd}, where the uneven illumination, haze, and out-of-focus blur degradation are presented in (a), (b), and (c), respectively. It can be observed that the performance of blood vessel segmentation only collapses on out-of-focus blurred fundus images. 

Traditional fundus image restoration methods~\cite{foracchia2005luminosity,hwang2012context} are mainly based on handcrafted priors. However, these model-based methods achieve unsatisfactory performance and generality due to the poor representing capacity. Recently, deep Convolutional Neural Networks (CNNs) have been widely used in natural image restoration and enhancement~\cite{gong2019autogan,isola2017image,zhu2017unpaired}, \emph{e.g.}, super resolution~\cite{ledig2017photo,wang2018esrgan,cai2021learning,sengupta2020desupgan,zhao2019data,hu2021pyramid,hu2020eg}, deraining~\cite{qian2018attentive}, deblurring~\cite{dbgan,mirnet,mprnet,hinet,mtrnn}, enlighten~\cite{jiang2021enlightengan,yang2020fidelity}, \emph{etc.} Inspired by the success of natural image restoration, CNNs have also been applied to fundus image restoration~\cite{williams2017fast,zhao2019data,sengupta2020desupgan,shen2020modeling,cheng2021secret,mahapatra2019image,luo2020dehaze,wang2021retinal,zhang2022double,raj2022novel}. 
Although impressive results have been achieved, CNN-based methods show limitations in capturing long-range dependencies. In recent years, the natural language processing (NLP) model, Transformer~\cite{vaswani2017attention} has been introduced into computer vision and outperformed CNN-based methods in many tasks. The Multi-head Self-Attention (MSA) in Transformer excels at modeling  non-local similarity and long-range dependencies. This advantage of Transformer may provide a possibility to address the limitations of CNN-based methods.

Existing deep learning methods rely on a large amount of LQ and HQ fundus image pairs. Unfortunately, real clinical benchmark has not been explored for fundus image reconstruction. There remains a data-hungry problem. As shown in Fig.~\ref{dataset} (b), to get more image pairs,  artificially designed degradation models such as Gaussian filter are used to synthesize degraded fundus images from their high-quality counterparts. However, as depicted in Fig.~\ref{dataset} (c), artificial degradation is fundamentally different from clinical degradation. As shown in Fig.~\ref{dataset} (d), the two CNN-based methods~\cite{cheng2021secret,shen2020modeling} trained with synthesized data fail in real fundus image restoration. 

In this paper, we investigate the real fundus image restoration problem, which has not been studied in the literature. Our work is the first attempt. To begin with, we establish a clinical benchmark, Real Fundus (RF), including 120 LQ and HQ real clinical fundus image pairs to alleviate the data-hungry issue. Based on this dataset, we propose a novel method, namely Transformer-based Generative Adversarial Network (RFormer), for real fundus image restoration. Specifically, the generator and discriminator are built up by the basic unit, Window-based Self-Attention Blocks (WSABs). The self-attention mechanism equipped with each basic block excels at capturing the non-local self-similarity and long-range dependencies, which are the main limitations of existing CNN-based methods. In particular, the generator adopts a U-shape structure to aggregate multi-resolution contextual information. Unlike previous CNN-based Generative Adversarial Networks (GANs), we adopt a Transformer-based discriminator to extract non-local image prior information and thus improve the ability of discriminator to distinguish restored fundus images from the ground-truth HQ fundus images. Our Transformer-based adversarial training scheme encourages the generator to create more plausible-looking natural and visually-pleasant images with more detailed contents and structural textures.  

Our contributions can be summarized as follows:
\begin{itemize}
\item We establish a new clinical  benchmark, RF, to evaluate algorithms in real fundus image restoration. To the best of our knowledge, this is the first real fundus image dataset.

\item We propose a novel Transformer-based method, RFormer, for real fundus image restoration. To the best of our knowledge, it is the first attempt to explore the potential of Transformer for this task in the literature. 

\item Comprehensive quantitative and qualitative experiments demonstrate that our RFormer significantly outperforms SOTA algorithms. Extensive experiments of downstream tasks further validate the effectiveness of our method.
\end{itemize}

\section{related work}
\label{sec:related work}

\subsection{Fundus Image Restoration}
\label{ssec:FIR}

Traditional fundus image restoration and enhancement methods~\cite{foracchia2005luminosity,hwang2012context} are mainly based on hand-crafted priors. For example, Setiawan $\textit{et al}.$~\cite{setiawan2013color} apply contrast limited adaptive histogram equalization (CLAHE) to fundus image enhancement. Some methods~\cite{ng2011total,wang2014nonlocal,fu2016weighted} decompose the reflection and illumination, achieving image enhancement and correction by estimating the solution in an alternate minimization scheme. However, these model-based methods achieve unsatisfactory performance and generality due to the poor representing capacity. With the development of deep learning, fundus image restoration has witnessed a significant progress.  CNNs~\cite{williams2017fast,zhao2019data,sengupta2020desupgan,shen2020modeling,cheng2021secret,mahapatra2019image,luo2020dehaze,wang2021retinal,zhang2022double,raj2022novel} apply a powerful learning model to restore LQ fundus images. For instance, Zhao $\textit{et al}.$~\cite{zhao2019data} propose an end-to-end deep CNN to remove the lesions on the fundus images of cataract patients. However, the cataract lesions are not caused by clinical fundus imaging. Sourya $\textit{et al}.$~\cite{sengupta2020desupgan} , Shen $\textit{et al}.$~\cite{shen2020modeling}, and Raj $\textit{et al}.$~\cite{raj2022novel} customize different synthetic degradation models to better simulate the degradation types in actual clinical practice. However, real fundus image degradation is more sophisticated than synthesized degradation. It is hard to simulate real degradation by artificial degradation models completely. Thus, models trained on synthesized data easily fail in real fundus image restoration. In addition, the CNN-based methods show limitations in capturing non-local self-similarity and long-rang dependencies, which are critical for fundus image reconstruction. 

\subsection{Generative Adversarial Network}
\label{ssec:gan}

Generative Adversarial Network (GAN) is firstly introduced in ~\cite{goodfellow2014generative} and has been proven successful in image synthesis~\cite{gong2019autogan,isola2017image,zhu2017unpaired}, and translation~\cite{isola2017image,zhu2017unpaired}. Subsequently, GAN is applied to image restoration and enhancement, \emph{e.g.}, super resolution~\cite{ledig2017photo,wang2018esrgan,cai2021learning,sengupta2020desupgan,zhao2019data,hu2021pyramid,hu2020eg}, deraining~\cite{qian2018attentive}, deblurring~\cite{dbgan}, enlighten~\cite{jiang2021enlightengan,yang2020fidelity}, dehazing~\cite{li2018single,ijcai2021-101}, image inpainting~\cite{yu2018generative,zheng2019pluralistic}, style transfer~\cite{zhu2017unpaired,li2016combining}, image editing~\cite{yang2019controllable,yang2020deep}, medical image enhancement~\cite{ma2021structure,chen2022novel,zhao2019data,sengupta2020desupgan}, and mobile photo enhancement~\cite{yuan2018unsupervised,chen2018deep}. Although GAN is widely applied in low-level vision tasks, few works are dedicated to improving the underlying framework of GAN, such as replacing the traditional CNN framework with Transformer. Jiang $\textit{et al}.$~\cite{jiang2021transgan} propose the first Transformer-based GAN, TransGAN, for image generation. Nonetheless, to the best of our knowledge, the Transformer-based GAN has not been involved in fundus image restoration. 

\subsection{Vision Transformer}
\label{ssec:transformer}
Transformer is proposed by~\cite{vaswani2017attention} for machine translation. Recently, Transformer has achieved great success in high-level vision, such as image classification~\cite{arnab2021vivit,dosovitskiy2020image,el2021xcit,liu2021swin,wu2020visual},  semantic segmentation~\cite{cao2021swin,liu2021swin,wu2020visual,zheng2021rethinking}, human pose estimation~\cite{li2021pose,li2021tokenpose,yang2021transpose,cai2020learning,luo2021efficient,cai2019res}, object detection~\cite{carion2020end,dai2021dynamic,liu2021swin,zhu2020deformable,huang2020joint}, $\textit{etc}.$ Due to the advantage of capturing long-range dependencies and  excellent performance in many high-level vision tasks, Transformer has also been introduced into low-level vision~\cite{restormer,cao2021video,chen2021pre,liang2021swinir,wang2021uformer,cai2022mst++,cai2022degradation}. SwinIR~\cite{liang2021swinir} uses Swin Transformer~\cite{liu2021swin} blocks to build up a residual network and achieve SOTA results in natural image restoration. Chen $\textit{et al}.$~\cite{chen2021pre} propose a large model IPT pre-trained on large-scale datasets with a multitask learning scheme. MST~\cite{mst} presents a spectral-wise Transformer for HSI reconstruction. Although Transformer has achieved impressive results in many tasks, its potential in fundus image restoration remains under-explored.

\begin{figure*}[!t]
\centerline{\includegraphics[width=2.05\columnwidth]{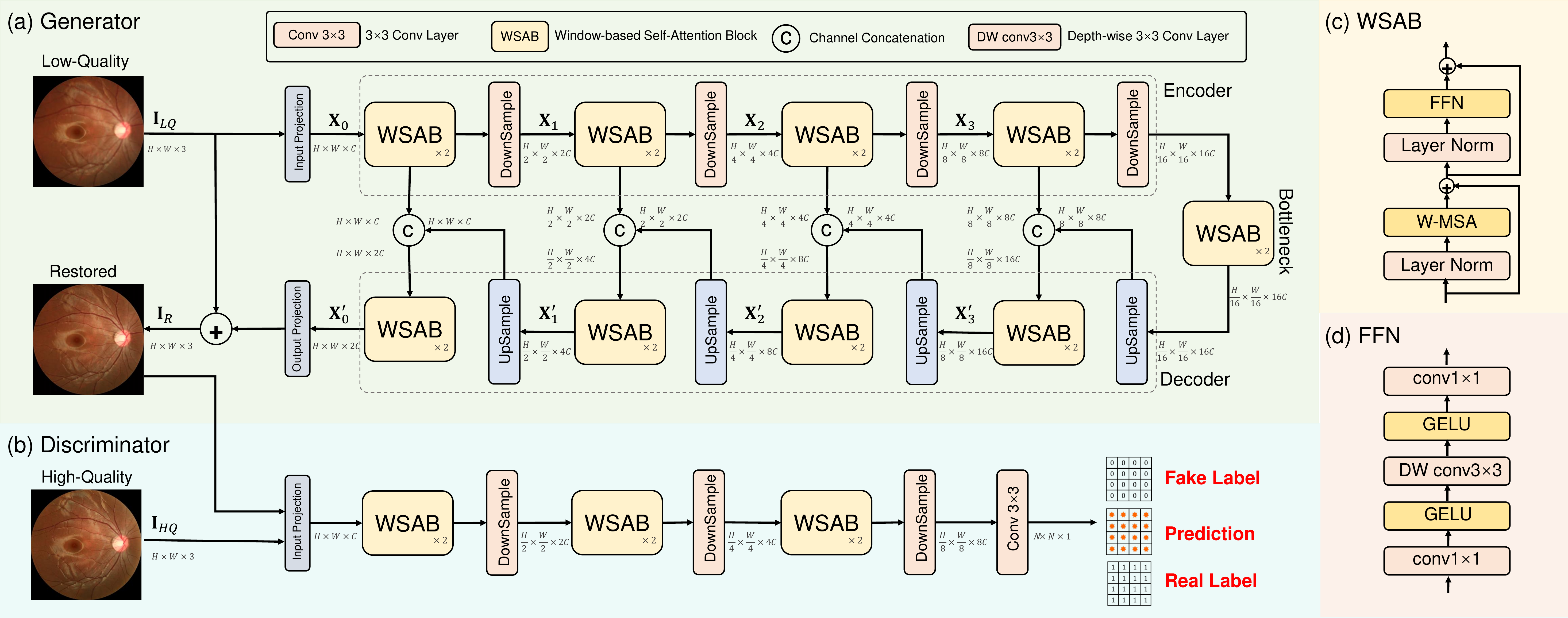}}
\caption{ \textbf{Architecture of our RFormer.} (a) The generator adopts a U-shaped structure, including an encoder, a bottleneck, and a decoder. (b) Different from CNN-based GANs, our discriminator is Transformer-based. (c) The basic unit of our RFormer is  Window-based Multi-head Self-Attention Block. (d) Feed-Forward Network consits of two 1$\times$1 $conv$ layers, two GLEU activations, and a depth-wise 3$\times$3 $conv$ layer.
}
\label{pipline}
\end{figure*}

\section{Methodology}
\label{sec:method}

\subsection{RFormer Architecture}
\label{ssec:modelstructure}
The architecture of RFormer is shown in Fig.~\ref{pipline}, where (a) and (b) depict the generator and discriminator. Fig.~\ref{pipline} (c) illustrates the proposed Window-based Self-Attention  Blocks (WSABs), which consists of a Feed-Forward Network (FFN) (detailed in Fig.~\ref{pipline} (d)), a Window-based Multi-head Self-Attention (W-MSA), and two layer normalization. 

The generator adopts a U-shaped~\cite{ronneberger2015u} architecture including an encoder, a bottleneck, and a decoder. The input LQ image is denoted as $\mathbf{I}_{LQ}\in\mathbb{R}^{{H \times W \times 3}}$. \textbf{Firstly}, the generator exploits a projection layer consisting of a $3 \times 3$ convolution (\emph{conv}) and LeakyReLU to extract shallow feature $\mathbf{I}_0\in\mathbb{R}^{{H \times W \times C}}$. \textbf{Secondly}, $4$ encoder stages are used for deep feature extraction on $\mathbf{I}_0$. Each stage is composed of two consecutive WSABs and one downsampling layer. We adopt a $4\times4$ \emph{conv} with stride 2 as the downsampling layer to downscale the spatial size of feature maps and double the channel dimension. Thus, the feature of the $i$-th stage in the encoder is denoted as $\mathbf{X}_i \in \mathbb{R}^{\frac{H}{2^i} \times \frac{H}{2^i} \times 2^{i}C}$. Here, $i = 0, 1, 2, 3$ indicates the four stages. \textbf{Thirdly}, $\mathbf{X}_3$ undergoes the bottleneck that consists of two WSABs. \textbf{Subsequently}, following the spirits of U-Net, we customize a symmetrical decoder, which also contains $4$ stages. Each stage of the decoder is also composed of two WSABs and one upsampling layer. Similarly, the feature maps of the $i$-th stage in the decoder is denoted as $\mathbf{X}'_i \in \mathbb{R}^{\frac{H}{2^i} \times \frac{H}{2^i} \times 2^{i+1}C}$. The upsampling layer is a bilinear interpolation followed by a 3$\times$3 \emph{conv} layer. To alleviate the information loss caused by downsampling in the encoder, skip connections are used for feature fusion between the encoder and decoder. \textbf{Finally}, after undergoing the decoder, the feature maps pass through a 3$\times$3 \emph{conv} layer to generate a residual image $\mathbf{I}^{'}\in\mathbb{R}^{{H\times W \times 3}}$. The restored fundus image can be obtained by $\mathbf{I}_{R} = \mathbf{I}_{LQ} + \mathbf{I}^{'}$.

\begin{figure}[!t]
\centerline{\includegraphics[width=\columnwidth]{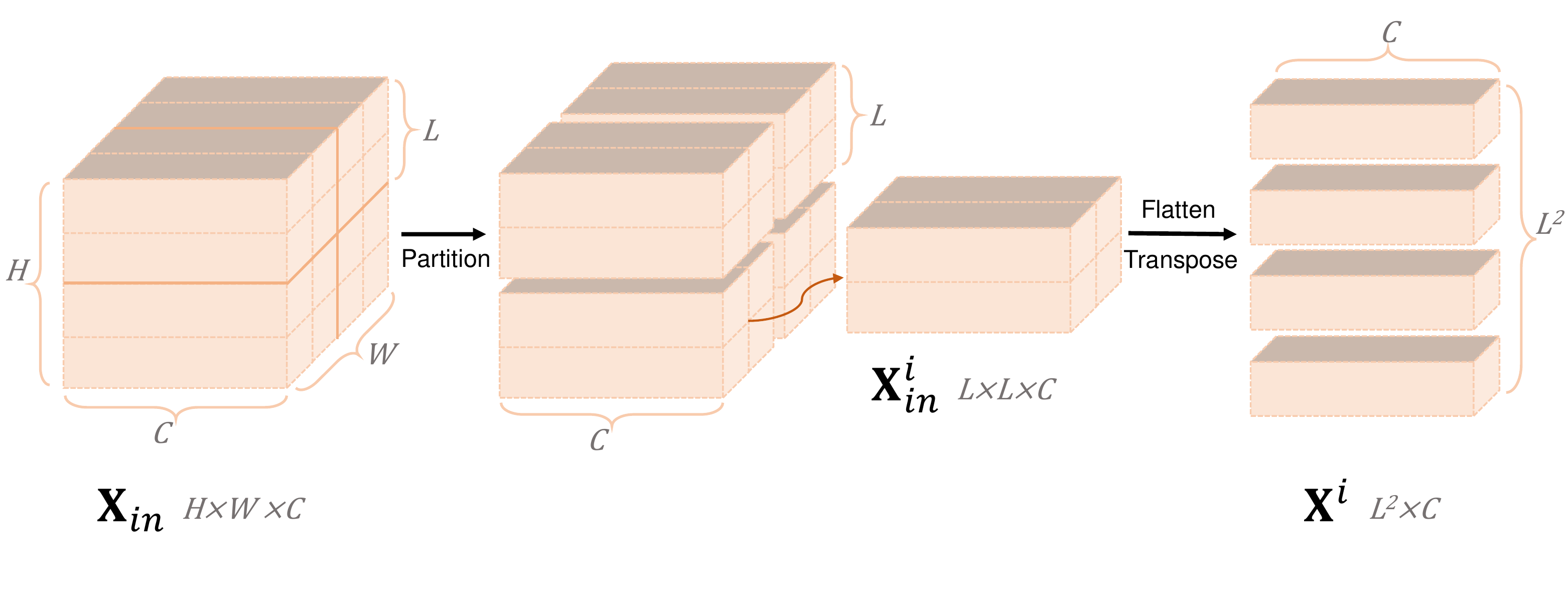}}
\caption{  \textbf{Illustration of the feature map partition.}
The feature maps are partitioned into non-overlapping windows, where the window-based multi-head self-attention (W-MSA) is conducted. }
\label{window}
\end{figure}

As shown in Fig.~\ref{pipline} (b), the discriminator aims to distinguish the restored fundus images from ground-truth high-quality fundus images. As analyzed in~\cite{isola2017image}, patch-level GAN is more effective than image-level GAN in capturing high-resolution and fine-grained image information, which is critical for restoring clinical fundus image restoration.  
Hence, we follow the adversarial training scheme based on image patches as PatchGAN and further propose a Transformer-based discriminator. More specifically, the discriminator employs the same architecture as the encoder in the generator followed by a 3$\times$3 \emph{conv} layer. The restored fundus image $\mathbf{I}_{R}\in\mathbb{R}^{{H\times W \times 3}}$ concatenated with the ground-truth HQ fundus image $\mathbf{I}_{HQ}\in\mathbb{R}^{{H\times W \times 3}}$ undergoes our proposed Transformer-based discriminator to generate the predicted map $\mathbf{F}\in\mathbb{R}^{{N\times N \times 1}}$. 

\subsection{Window-based Self-Attention Block}
\label{ssec:PTB}
The emergence of Transformer provides an alternative to address the limitations of CNN-based methods in modeling non-local self-similarity and long-range dependencies. However, as analyzed in Swin Transformer~\cite{liu2021swin}, the computational cost of the standard global Transformer is quadratic to the spatial size of the input feature ($HW$). This burden is nontrivial and sometimes unaffordable. To tackle this problem, we adopt the Window-based Multi-head Self-Attention (W-MSA) ~\cite{liu2021swin} as the self-attention mechanism and integrate it with the basic Transformer unit. The computational complexity of W-MSA is linear to the spatial size, which is much cheaper than that of standard global MSA. Inspired by Swin Transformer~\cite{liu2021swin}, we add window shift operations(WSO) in our proposed Window-based Multi-head Self-Attention Block (WSAB) to introduce cross-window connections. The components of our proposed WSAB are shown in Fig.~\ref{pipline} (c). WSAB consists of a W-MSA, an FFN, and two layer normalization. The details of FFN are shown in Fig.~\ref{pipline} (d). Then WSAB can be formulated as 
\begin{small}
\begin{equation}
    \begin{aligned}
   & \mathbf{F}^{\prime} = \text{W-MSA}(\text{LN}(\mathbf{F}_{in}))+\mathbf{F}_{in},  \\
   & \mathbf{F}_{out} = \text{FFN}(\text{LN}(\mathbf{F}^{\prime}))+\mathbf{F}^{\prime},
    \end{aligned}
\end{equation}
\end{small}
where $\mathbf{F}_{in}$ represents the input feature maps of a WSAB. $\text{LN}(\cdot)$ represents the layer normalization. $\mathbf{F}^{\prime}$ and $ \mathbf{F}_{out}$ denote the output feature of W-MSA and FFN  respectively.
\subsubsection{Window-based Multi-head Self-Attention}
\label{sssec:W-MSA}
Instead of using global correspondence, we partition the feature map into  non-overlapping windows. Subsequently, the token interactions are calculated inside each window. As shown in Fig.~\ref{window}, given the input feature map $\mathbf{X}_{in}\in\mathbb{R}^{{H \times W \times C}}$ with ${{H}}$ and ${{W}}$ being the height and the width. We partition $\mathbf{X}_{in}$ into $L \times L$ non-overlapping windows. The feature of the $i$-th window is denoted as  $\mathbf{X}_{in}^{i}\in\mathbb{R}^{{L \times L \times C }}$, where  $i \in \{1,2,\cdots, N\}$ and $N = H W / {L^{2}}$. $\mathbf{X}_{in}^{i}$ is flattened and transposed into $\mathbf{X}^{i}\in\mathbb{R}^{{ L^{2}\times C}}$. We conduct MSA on  $\mathbf{X}^{i}$. Firstly,  $\mathbf{X}^{i}$ is linearly projected into \emph{query} $\mathbf{Q}^{i}$, \emph{key} $\mathbf{K}^{i}$, and \emph{value} $\mathbf{V}^{i} \in \mathbb{R}^{L^{2} \times C}$:


\begin{small}
\begin{equation}
    \begin{aligned}
   \mathbf{Q}^{i} = \mathbf{X}^{i} \mathbf{W}^{Q}, \mathbf{K}^{i} = \mathbf{X}^{i} \mathbf{W}^{K}, \mathbf{V}^{i} = \mathbf{X}^{i} W^{V},
    \end{aligned}
\end{equation}
\end{small}
where $\mathbf{W}^{Q},\mathbf{W}^{K},\mathbf{W}^{V} \in\mathbb{R}^{\bf{C\times C}} $  are learnable parameters, denoting the projection matrices of the \emph{query}, \emph{key} and \emph{value}. We respectively split $\mathbf{Q}^{i}$, $\mathbf{K}^{i}$ and $\mathbf{V}^{i}$ into $k$ heads along the channel dimension:  $\mathbf{Q}^{i} = [\mathbf{Q}^{i}_1,\ldots,\mathbf{Q}^{i}_k]$, $\mathbf{K}^{i} = [\mathbf{K}^{i}_1,\ldots,\mathbf{K}^{i}_k]$ and $\mathbf{V}^{i} = [\mathbf{V}^{i}_1,\ldots,\mathbf{V}^{i}_k]$. The dimension of each head is $d_{k} = C / k$. The Self-Attention (SA) for head $j$ is formulated as
\begin{small}
\begin{equation}
    \text{SA}(\mathbf{Q}^i_j,\mathbf{K}^i_j,\mathbf{V}^i_j) = \text{softmax}(\frac{\mathbf{Q}^i_j {\mathbf{K}^i_j}^T}{\sqrt{d_k}})\mathbf{V}^i_j,
\end{equation}
\end{small}
where $\mathbf{Q}^i_j,\mathbf{K}^i_j$ and $\mathbf{V}^i_j$ respectively represent the  \emph{query}, \emph{key} and \emph{value} of head $j$ respectively. The output tokens  $\mathbf{X}^i_{o} \in \mathbb{R}^{L^2 \times C}$ of the $i$-th window can be obtained by
\begin{small}
\begin{equation}
    \mathbf{X}^i_{o} = \mathop{\text{Concat}}\limits_{j=1}^{k}\big(\text{SA}(\mathbf{Q}^{i}_j,\mathbf{K}^{i}_j,\mathbf{V}^{i}_j)\big)\mathbf{W}^{O} + \mathbf{B}, 
\end{equation}
\end{small}

where Concat$(\cdot)$ denotes the concatenating operation, $\mathbf{B} \in \mathbb{R}^{L^2\times C}$ represents the position embedding, and  $\mathbf{W}^{O}\in\mathbb{R}^{{C \times  C}} $ are learnable parameters. We reshape $\mathbf{X}^{i}_{o}$ to obtain the output window feature map $\mathbf{X}^{i}_{out} \in\mathbb{R}^{{L \times L \times C }}.$ Finally, we merge all the patch representations $\{\mathbf{X}_{out}^{1}, \mathbf{X}_{out}^{2},\mathbf{X}_{out}^{3}, \cdots , \mathbf{X}_{out}^{N} \}$ to obtain the output feature maps $\mathbf{X}_{out} \in\mathbb{R}^{{H \times W \times C }}$.
\subsubsection{Feed-Forward Network}
\label{sssec:FFN}
As depicted in Fig.~\ref{pipline} (d), the Feed-Forward Network (FFN) consists of a $1 \times 1$ $conv$ layer with a GELU activation, a depth-wise $3 \times 3$ $conv$ layer with a GELU activation, and another $1 \times 1$ $conv$ layer. 

\subsection{Loss Functions}
\label{ssec:loss functions}
During the training procedure, we exploit the weighted sum of four loss functions as the overall  training objective. They are described and analyzed in the following part.

\subsubsection{Charbonnier Loss}
\label{sssec:charbonnier-loss}
The first loss function is the Charbonnier loss between the restored and ground-truth HQ images:
\begin{small}
\begin{equation}
     \mathcal{L}_{1}(\mathbf{I}_R , \mathbf{I}_{HQ}) = \sqrt{ \left \| \mathbf{I}_R - \mathbf{I}_{HQ} \right \|^2 + \varepsilon^2 }
\end{equation}
\end{small}
where $\mathbf{I}_R$ denotes the restored fundus image, $\mathbf{I}_{HQ}$ represents the ground-truth HQ fundus image, and $\varepsilon$ denotes a constant which is empirically set to $10^{-3}$ for all the experiments.

\subsubsection{Fundus Quality Perception Loss}
\label{sssec:perceptual-loss}
Unlike natural images, fundus images have specific acquisition process and anatomical structures. This indicates fundus images have highly similar styles. Therefore, we exploit high-level feature constraints to improve the perceptual quality and  encourage the network to capture the fundus anatomical structures and styles. To this end, we propose Fundus Quality Perception Loss (FQPLoss). More specifically, we adopt VGG-19~\cite{simonyan2014very} as the perception network and pre-train it on the fundus image quality evaluation dataset, Eye-Q~\cite{fu2019evaluation} with the fundus image quality classification task. The Eye-Q dataset has 28,792 fundus images with three-level quality grading. The perception network trained on the Eye-Q dataset is capable of extracting the difference of high-level features between different qualities of fundus images. Subsequently, our FQPLoss can be formulated as 
\begin{small}
\begin{equation}
     \mathcal{L}_{fqp} = \frac{1}{HW} \sum_{i = 1}^{H} \sum_{j = 1}^{W}(\phi(\mathbf{I}_{R})(i,j) - \phi(\mathbf{I}_{HQ})(i,j))^2
\end{equation}
\end{small}
where $H$ and $W$ denote the height and width of the fundus image. $\phi(\cdot)$ denotes the feature extraction function of the pre-trained perception network. Our FQPLoss is customized to assess a solution with respect to perceptually relevant characteristics. By minimizing the FQPLoss $\mathcal{L}_{fqp}$, the model is encouraged to capture more high-level discriminative features and generate more visually-pleasant results.

\subsubsection{Adversarial  Loss}
\label{sssec:adversarial-loss}
Our proposed Transformer-based discriminator aims to distinguish the restored fundus images from ground-truth HQ fundus images. More specifically, the discriminator outputs a patch score map $\mathbf{F}\in\mathbb{R}^{{N_H \times N_W \times 1}}$, where $N_H$ and $N_W$ denote the height and width. The score of each position indicates how realistic the corresponding fundus image patch is. Then the adversarial loss is formulated as
\begin{small}
\begin{equation}
    \begin{aligned}
   & \mathcal{L}_{adv}^{D} =  \frac{1}{N_H\times N_W}\sum_{i = 1}^{N_H\times N_W} (D(\mathbf{I}_R)[i])^2 + (D(\mathbf{I}_{HQ})[i]-1)^2, \\
   & \mathcal{L}_{adv}^{G} =  \frac{1}{N_H\times N_W} \sum_{i = 1}^{N_H\times N_W} (D(\mathbf{I}_R)[i]-1)^2,
    \end{aligned}
\end{equation}
\end{small}
where $D(\cdot)$ denotes the mapping function of our proposed Transformer-based discriminator. $\mathcal{L}_{adv}^{G}$ trains the generator to fool the discriminator by generating more realistic restored fundus images. In contrast, $\mathcal{L}_{adv}^{D}$ encourages the discriminator to distinguish the restored images from real images.

\subsubsection{Edge Loss}
\label{sssec:Edge loss}
To enhance the high-frequency edge details, we exploit the edge loss function that focuses on the gradient information of images and enhances edge textures. To be specific, the edge loss function is formulated as
\begin{small}
\begin{equation}
     \mathcal{L}_{edge}(\mathbf{I}_R , \mathbf{I}_{HQ}) = \sqrt{ \left \| \Delta(\mathbf{I}_R) - \Delta(\mathbf{I}_{HQ}) \right \|^2 + \varepsilon^2 },
\end{equation}
\end{small}
where $\Delta(\cdot)$ represents the Laplacian operator. 

\subsubsection{The Overall Loss Function}
Finally, the overall training objective is the weighted sum of the above four loss functions:
\begin{small}
\begin{equation}
     \mathcal{L} = \mathcal{L}_1 + \lambda_{1} \mathcal{L}_{fqp} + \lambda_{2} \mathcal{L}_{edge} + \lambda_{3} (\mathcal{L}_{adv}^{G} + \mathcal{L}_{adv}^{D}),
\end{equation}
\end{small}
where $\lambda_{1}, \lambda_{2}, \lambda_{3}$ are three hyper-parameters controlling the importance balance of different loss functions. Our proposed RFormer is end-to-end trained by minimizing $\mathcal{L}$. The weights of the perception network are fixed. Each mini-batch training procedure can be divided into two steps: (i) Fix the discriminator and train the generator. (ii) Fix the generator and train the discriminator. This adversarial training scheme encourages the reconstructed fundus images to be more photo-realistic and closer to the real clinical HQ fundus image manifold.

\section{Real Fundus}
\label{rf}
This section introduces our clinical benchmark, Real Fundus (RF). It consists of 120 LQ and HQ clinical fundus image pairs with the spatial size of $2560 \times 2560$. The training and testing subsets are split in proportional to 3:1. Since blur significantly impacts clinical diagnosis and automated image analyzing systems, it is set to the primary degradation type of LQ fundus images. Besides, there are other degradation types such as artifacts and uneven illumination which are inevitably introduced in the fundus image capturing process.

\begin{figure*}[!t]
\centerline{\includegraphics[width=2\columnwidth]{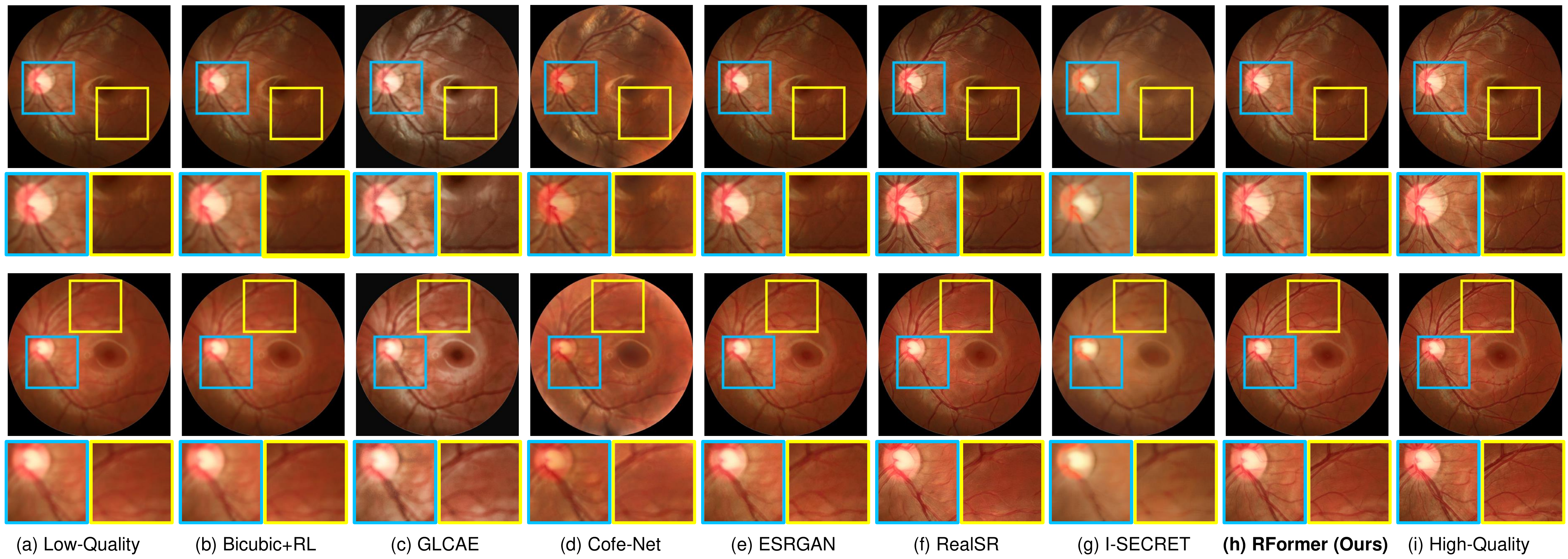}}
\caption{ \textbf{Restored fundus image comparisons on our RF.} Six SOTA methods and our proposed RFormer are included. (a) LQ fundus image. (b) Bicubic+RL~\cite{tai2010richardson}. (c) GLCAE~\cite{tian2017global}. (d)  Cofe-Net~\cite{shen2020modeling}. (e) ESRGAN~\cite{wang2018esrgan}. (f) RealSR~\cite{ji2020real}. (g) I-SECRET~\cite{cheng2021secret}. (h) Our proposed RFormer. (i) Ground-truth HQ fundus image. Our RFormer reconstructs more detailed contents and structural testures.}
\label{duibi1}
\end{figure*}


\subsection{Data Collecting Process}
\label{ssec:sources}

The collection process of our RF obtains the exemption determination from Shenzhen Eye Hospital and contains three steps:  capturing,  selecting, and calibrating fundus images. 

\subsubsection{Capturing}
\label{sssec:Capturing}
Instead of exploiting artificial degradation models (\emph{e.g.}, Gaussian Filter.) to synthesize LQ fundus images as shown in Fig.~\ref{dataset} (b), we directly use the degraded fundus images from the fail cases in practical capturing. As depicted in Fig.~\ref{dataset} (a), the fundus images are captured by ophthalmologists using a ZEISS VISUCAM200 fundus camera, which is a mainstream product of fundus camera. The price of ZEISS VISUCAM200 fundus camera is about 350,000 RMB. 
We select clinical fundus images from patients of different ages and different fundus states (\emph{e.g.}, leopard fundus, hemorrhage, microaneurysms, and drusen ) to expand the scope of our RF.

\subsubsection{Selecting}
\label{sssec:Selecting}
 When LQ fundus images are captured in practice, the operator will repeat capturing until HQ fundus images are obtained. Subsequently, we manually select LQ and HQ fundus image pairs of the same eye. To ensure the diversity of RF and avoid similar data, only one image pair is selected with one eye. Note that only HQ clear fundus images captured by experienced ophthalmologists can be used as the ground truths of degraded LQ images. Based on these strict criteria, we finally select 120 LQ and HQ fundus image pairs from the eye hospital database containing more than 30,00 eye instances. Each instance contains multiple fundus images.
 
 \subsubsection{Calibrating} 
 \label{sssec:Calibrating}
 After selecting fundus image pairs, we observe two issues in raw unprocessed fundus data. Firstly, the LQ and HQ fundus images are spatially misaligned (as illustrated in Fig.~\ref{dataset}(a)). Secondly, there is a large black area around the eyeball. This black area is uninformative and may easily degrade the performance of the restoration model during the training procedure. Thus, to improve the quality of our RF, we calibrate the collected dataset using the software, Photoshop. Specifically, we first spatially align the image pairs and then cut off the black area around the eyeball. 
 
\begin{figure}[!t]
\centerline{\includegraphics[width=\columnwidth]{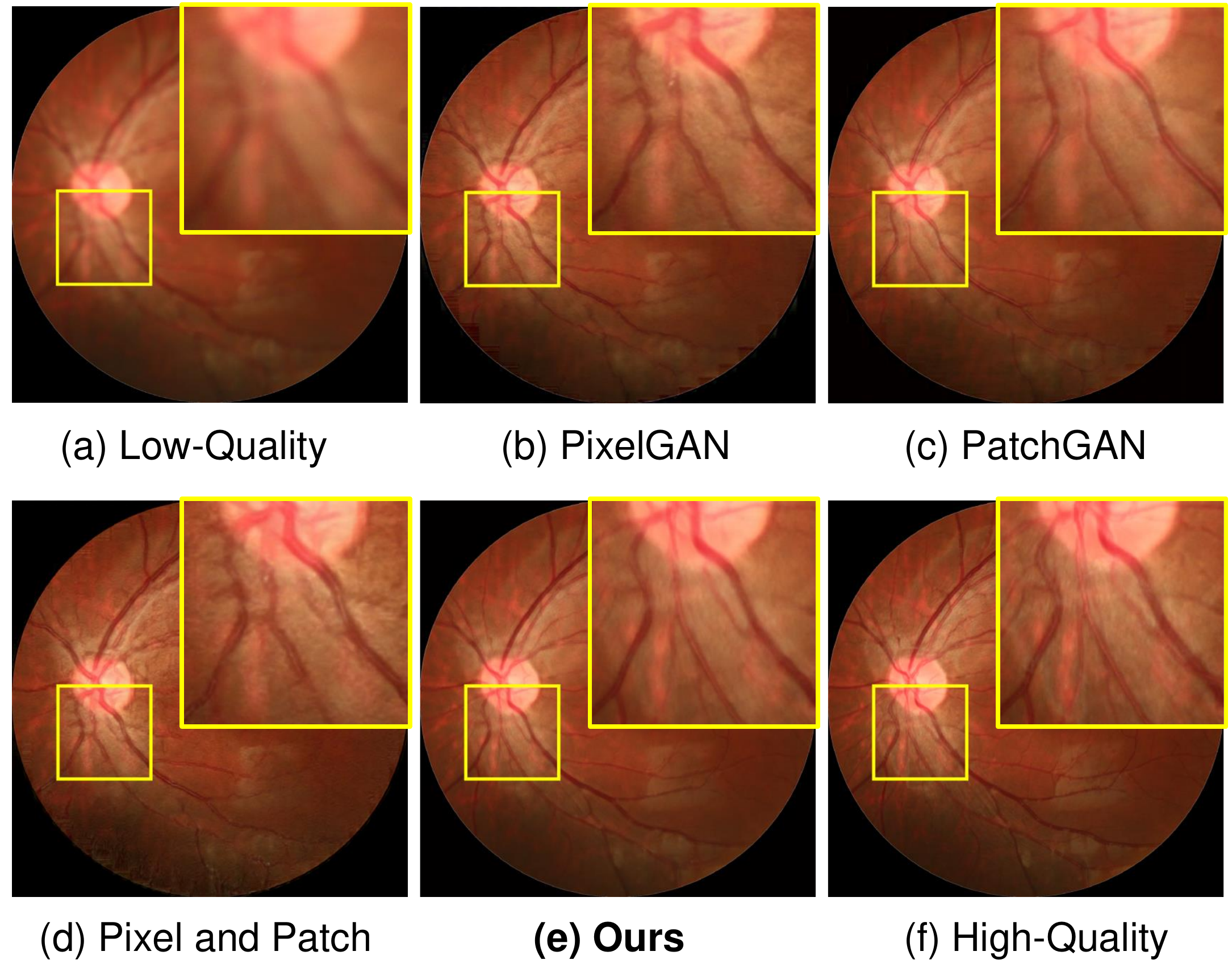}}
\caption{ \textbf{CNN-based \emph{vs.} Transformer-based discriminators.}
(a) LQ fundus image. (b) Using CNN-based PixelGAN~\cite{isola2017image}. (c) Using CNN-based  PatchGAN~\cite{isola2017image}. (d) Using both CNN-based PixelGAN and PatchGAN. (e) Equipped with our proposed Transformer-based discriminator. (f) Ground truth HQ fundus image. Our Transformer-based discriminator significantly surpasses traditional CNN-based discriminators in terms of recovering detailed contents and preserving the anatomical structure.}
\label{gan}
\end{figure}

\begin{figure}[!t]
\centerline{\includegraphics[width=\columnwidth]{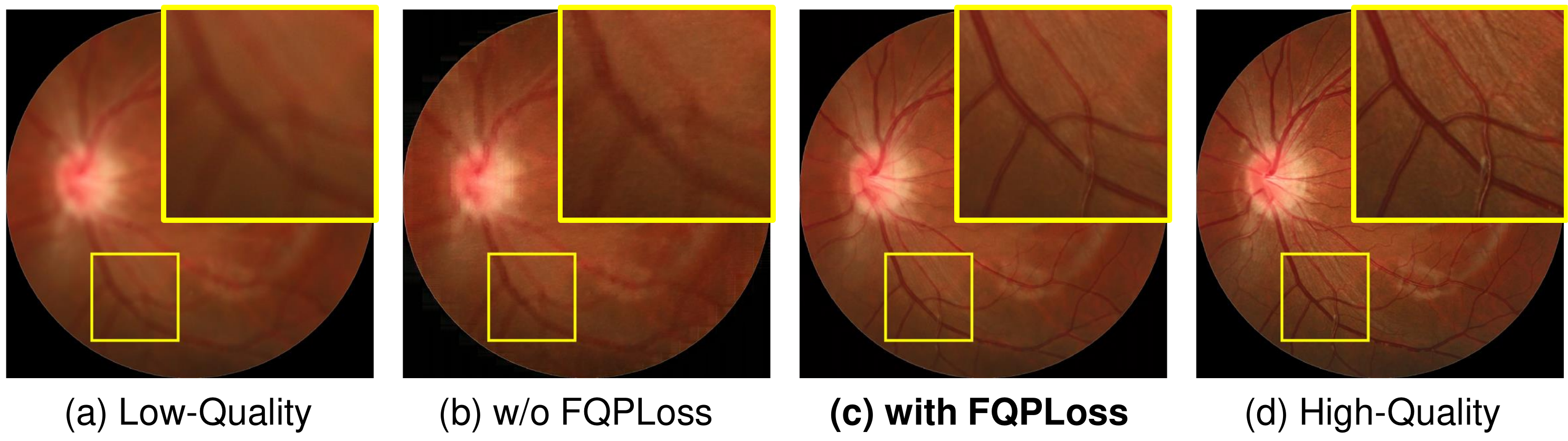}}
\caption{ \textbf{Ablation study of our FQPLoss.} 
(a) shows the LQ fundus image. (b) and (c) depict the images restored by RFormers without and with using FQPLoss, respectively. (d) illustrates the ground-truth HQ fundus image. With our FQPLoss, the model restores more detailed anatomical structure contents and high-frequency textures. }
\label{fqploss}
\end{figure} 
 
\subsection{Comparisons with Synthetic Dataset }
\label{ssec:csd}
\subsubsection{LQ Fundus Images} We compare the LQ images from our RF and synthetic dataset in Fig.~\ref{dataset} (c). As can be seen from the zoom-in patches that the artificially synthesized degradation is fundamentally different from the real clinical degradation.

\subsubsection{Domain Discrepancy} To validate the huge domain discrepancy between the synthetic and real clinical datasets\tr{,} we adopt two CNN-based fundus image restoration methods, I-SECRET~\cite{cheng2021secret} and Cofe-Net~\cite{shen2020modeling}, to conduct ablation study. We train them with the synthetic data and then test them on our RF. As shown in Fig.~\ref{dataset} (d), the two  models fail to reconstruct the real clinical LQ fundus images. They either yield over-smooth results sacrificing detailed contents, or introduce visually unpleasant artifacts. Since the synthetic data can not be applied to real fundus image restoration, it still remains a severe data-hungry issue. To meet with this research requirement, we establish a large scale clinical dataset, RF. To the best of our knowledge, this is the first work contributing a real clinical fundus image restoration benchmark.


\begin{figure*}[!t]
\centerline{\includegraphics[width=2.05\columnwidth]{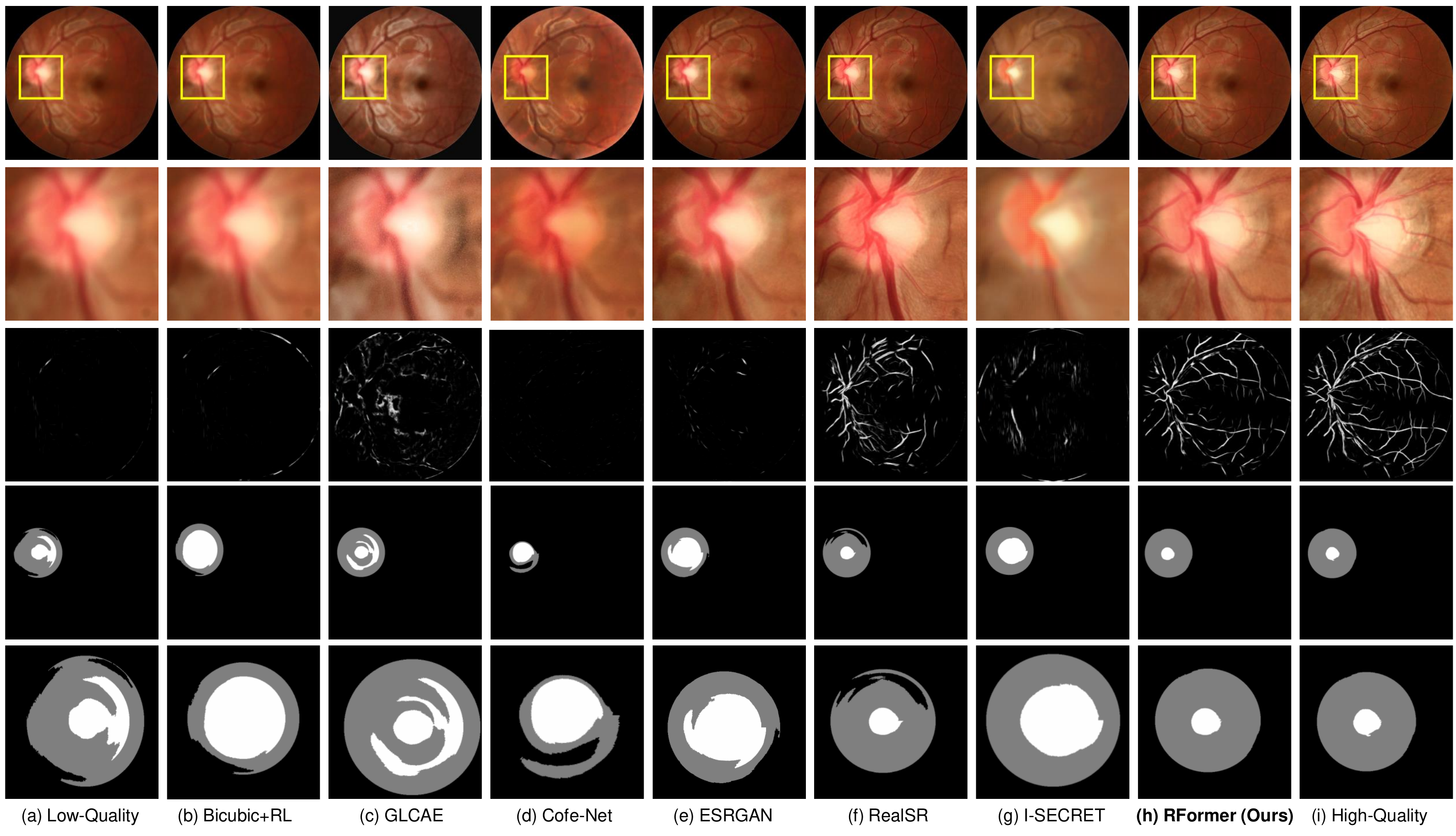}}
\caption{ \textbf{Vessel segmentation and optic disc/cup detection  on our RF}. From top to bottom are fundus images, optic disc/cup patches, vessel segmentation maps, optic disc/cup detection maps, and the zoom-in patches of the optic disc/cup detection maps.
(a) LQ fundus image.(b) Bicubic+RL~\cite{tai2010richardson}. (c) GLCAE~\cite{tian2017global}. (d)  Cofe-Net~\cite{shen2020modeling}. (e) ESRGAN~\cite{wang2018esrgan}. (f) RealSR~\cite{ji2020real}. (g) I-SECRET~\cite{cheng2021secret}. (h) Our RFormer. (i) HQ fundus image. Our restoration method can improve the performance of vessel segmentation and optic disc/cup detection.}
\label{fenge2}
\end{figure*}

\begin{figure*}
    \centering
	\includegraphics[scale=0.4]{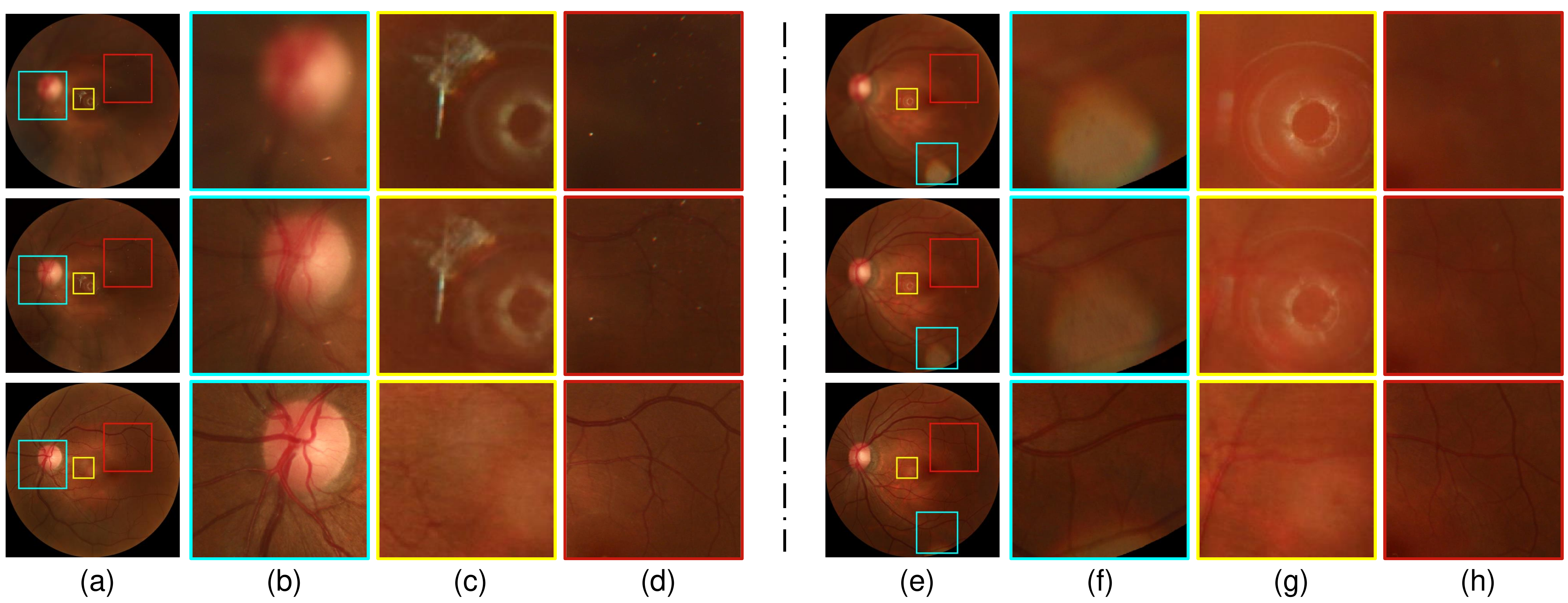}
	\caption{ {\bf Fail cases of Rformer on our RF.} In the (a) and (e) column, from top to bottom are low-quality fundus image, restored fundus image, and high-quality fundus image. (b), (c), and (d) are all the zoom-in patches of (a). (f), (g), and (h) are all the zoom-in patches of (e).}
    \label{fig3}
    \vspace{-3mm}
\end{figure*}

\section{Experiments}
\label{experiment}

\subsection{Implementation Details}
\label{ssec:details}
During the training procedure, fundus images are first cropped into the patches with the size of 128$\times$128. Then the patches are fed into our proposed RFormer. The Adam~\cite{adam} optimizer ($\beta_1$=0.9, $\beta_2$=0.999) is adopted. The initial learning rate is set to $1 \times 10^{-4}$.  The cosine annealing strategy~\cite{loshchilov2016sgdr} is employed to steadily decrease the learning rate from the initial value to $1 \times 10^{-6}$ during the training procedure. Our RFormer is implemented by PyTorch. It takes about 12h using an NVIDIA RTX 3090 GPU to train for 100 epochs. The mini-batch size is set to 4. Random flipping and rotation are used for data augmentation. In the testing phase, the input is the whole image with the size of 2560$\times$2560 for fair comparison with other methods. We adopt peak signal-to-noise ratio (PSNR) and structural similarity (SSIM)~\cite{ssim} as the metrics to evaluate the fundus image reconstruction performance. 

\begin{table}
	\footnotesize
	\centering	
	\caption{Quantitative comparisons with SOTA algorithms on our RF and cross-validation results of Rformer.}
	\scalebox{1.05}{
			\begin{tabular}{c c c c c}
				\toprule
				 Method & PSNR & SSIM  & Params(M) &FLOPS(G) \\
				\midrule
				Cofe-Net~\cite{shen2020modeling} & 17.26& 0.789 &39.31 & 22.48\\
				GLCAE~\cite{tian2017global} & 21.37 & 0.570 & -- & -- \\
				I-SECRET~\cite{cheng2021secret} & 24.57 & 0.854 & 10.85 & 14.21\\
				Bicubic+RL~\cite{tai2010richardson}  &25.34  & 0.824  & -- & -- \\
				ESRGAN~\cite{wang2018esrgan} &26.73  & 0.823   & 15.95 & 18.41 \\
				RealSR~\cite{ji2020real} &27.99 & 0.850  &15.92  & 29.42\\
				MST~\cite{mst} &28.13 & 0.854 & 3.48& 3.59\\
				\midrule
				\bf RFormer (Ours)    &  28.32 & \bf 0.873 &21.11 & 11.36 \\
				5-Fold Cross-valid & 28.15 & 0.848 &21.11 & 11.36\\
				10-Fold Cross-valid  & \bf 28.38 & 0.863 & 21.11 & 11.36 \\
				\bottomrule
	\end{tabular}}
	\label{tab:sota}
\end{table}

\begin{table}
	\footnotesize
	\centering	
	\caption{ Ablation study of the loss functions and window shift operations(WSO).}
	\scalebox{1.3}{\hspace{0mm}\begin{tabular}{c c c c c c }
				\toprule
				 {WSO}&{$\mathcal{L}_{1} $} &{$\mathcal{L}_{fqp} $}  &{$\mathcal{L}_{edge} $} &PSNR &SSIM  \\
				\midrule
				\checkmark&\checkmark& &  &27.42 &  0.840  \\
				\checkmark&\checkmark&\checkmark &  &27.90&  0.860   \\
				 & \checkmark& \checkmark& \checkmark & 28.20 &0.866  \\
    			\checkmark& &\checkmark & \checkmark &28.04&  0.858  \\
    			\checkmark&\checkmark& & \checkmark &27.25&  0.744  \\
    			\checkmark&\checkmark&\checkmark & \checkmark &\textbf{28.32}&\textbf{0.873}  \\
				\bottomrule
	\end{tabular}}
	\label{tab:loss}
\end{table}

\subsection{Comparisons with State-of-the-Art Methods}
\label{ssec:sota}

We provide quantitative comparisons between our RFormer with  seven SOTA methods including two model-based methods (GLCAE~\cite{tian2017global} and Bicubic+RL~\cite{tai2010richardson}), four CNN-based methods (RealSR~\cite{ji2020real}, ESRGAN~\cite{wang2018esrgan}, I-SECRET~\cite{cheng2021secret}, and Cofe-Net~\cite{shen2020modeling}), and one Transformer-based method (MST~\cite{mst}). The quantitative comparisons on our RF are shown in Table~\ref{tab:sota}, the proposed RFormer outperforms other competitors in terms of PSNR and SSIM. Specifically, RFormer achieves 0.33 and 1.59 dB improvement in PSNR when compared to RealSR~\cite{ji2020real} and ESRGAN~\cite{wang2018esrgan}. 

\begin{table}
	\footnotesize
	\centering	
	\caption{ CNN-based \emph{vs.} our Transformer-based discriminators.}
	\scalebox{0.98}{
			\begin{tabular}{c c c c c }
				\toprule
				 Discriminator &PatchGAN &PixelGAN &Pixel and Patch &\bf Ours\\
				\midrule
				PSNR &27.52 &28.00 &27.70 &\bf 28.32  \\
				SSIM &0.822 &0.790 &0.839 &\bf 0.873  \\
				\bottomrule
	\end{tabular}}
	\label{tab:d}
\end{table}

To verify the robustness of our RFormer, we conduct 5-fold and 10-fold cross-validation. The results are shown in Table~\ref{tab:sota}. As can be seen that our Rformer still achieves robust results, \emph{e.g.}, 28.15 dB in 5-fold cross-validation and 28.38 dB in 10-fold cross-validation. The small gap in performance suggests that the overfitting is moderate while the effectiveness of RFormer is reliable and promising.

\begin{table}
	\footnotesize
	\centering	
	\caption{Ablation study of the patch size of our  discriminator.}
	\scalebox{0.95}{
			\begin{tabular}{c c c c c c} 
				\toprule
				 Patch Size &$22 \times 22  $  &\bf 40 $\times$ 40 &$46  \times 46   $& $62 \times 62  $ &$128  \times 128   $\\
				\midrule
				PSNR &28.23 &\bf 28.32 &28.12 & 28.02& 28.03  \\
				SSIM &0.866 &\bf 0.873 &0.827&0.862  &0.796  \\
				\bottomrule
	\end{tabular}}
	\label{tab:patchsize}
\end{table}

Fig.~\ref{duibi1} depicts the  qualitative comparisons on RF. It can be observed that Bicubic+RL~\cite{tai2010richardson}, GLCAE~\cite{tian2017global}, I-SECRET~\cite{cheng2021secret}, and Cofe-Net~\cite{shen2020modeling} yield over-smooth results and fail to restore the LQ blurry fundus images. Although ESRGAN~\cite{wang2018esrgan} and RealSR~\cite{ji2020real} can reconstruct more high-frequency edge details, they do not maintain the authenticity and anatomical structure of the original  fundus. Some undesired artifacts are introduced to the restored images, which may severely mislead the clinical diagnosis. In contrast, our RFormer is capable of restoring more fine-grained contents and structural details without introducing artifacts. Thus, the fundus anatomical structure can be well preserved.

\begin{table*}
	\footnotesize
	\centering	
	\caption{ The results of image quality evaluation in user study.}
	\scalebox{1.1}{
	\begin{tabular}{c c c c c c c c}
    \toprule
     Method & GLCAE~\cite{tian2017global} & Bicubic+RL~\cite{tai2010richardson}  &Cofe-Net~\cite{shen2020modeling} &I-SECRET~\cite{cheng2021secret}  & ESRGAN~\cite{wang2018esrgan} & RealSR~\cite{ji2020real}  &\bf RFormer (Ours)\\
    \midrule
    Score &37.49 &31.00 &35.67&24.81&59.33&77.20 &\bf 83.21  \\
    \bottomrule
  \end{tabular}}
  \label{tab:user}
\end{table*}

\subsection{Ablation Study}
\label{ssec:ablation}

\subsubsection{FQPLoss}
We adopt RFormer as the restoration model to conduct an ablation study to validate the effect of our  FQPLoss. As listed in Table~\ref{tab:loss}, when the FQPLoss is applied, the PSNR and SSIM are increased by 1.07 dB and 0.129,  respectively. In addition, we provide visual comparisons in Fig.~\ref{fqploss}. As depicted in Fig.~\ref{fqploss} (b), the model yields an over-smooth fundus image and fails in reconstructing the fine-grained vessel details without FQPLoss. As shown in Fig.~\ref{fqploss} (c), when using our FQPLoss, the model restores more detailed anatomical structure contents and high-frequency textures.

\subsubsection{Discriminator}
We conduct ablation study to compare our proposed Transformer-based discriminator with Traditional CNN-based discriminators. Please note that the Transformer-based generator remains unchanged. The results are reported in Table~\ref{tab:d}. Compared with the discriminators in CNN-based PatchGAN~\cite{isola2017image}, PixelGAN~\cite{isola2017image}, $\textit{etc},$ our Transformer-based discriminator yields the best performance. We provide qualitative comparisons in Fig.~\ref{gan}. It can be observed that our Transformer-based discriminator significantly outperforms CNN-based discriminators in terms of recovering detailed contents and preserving the anatomical structure consistency.

\subsubsection{Patch Size}
We experimentally analyze the effect of the patch size set in the Transformer-based discriminator. The results are shown in Table~\ref{tab:patchsize}. Our RFormer achieves the best restoration result with the patch size of 40$\times$40. 

\subsubsection{Window Size}
We change the window size of W-MSA and conduct experiments to study its effect. The results are reported in Table~\ref{tab:window}. It can be observed that our RFormer yields the best result when the window size is set to 8$\times$8.  

\begin{table}
	\footnotesize
	\centering	
	\caption{ Ablation study of the window size of our proposed WSAB.}
	\scalebox{1.1}{
			\begin{tabular}{c c c c c }
				\toprule
				 Window Size &PSNR  &SSIM  &Params(M) &FLOPS(G)   \\
				\midrule
				w/o WSAB &14.27 &0.601 &34.53 &16.38 \\
				$4\times 4 $  & 28.23 &0.873 &\bf 20.67 &\bf 10.83 \\
				\bf 8 $\times$ 8 &\bf 28.32 & \bf 0.873&21.11&11.36 \\
				$16 \times 16  $ &27.82&0.819 & 27.54 &13.46\\
				$32 \times 32 $ & 27.63& 0.812& 47.21& 18.59\\
				\bottomrule
	\end{tabular}}
	\label{tab:window}
\end{table}

\subsubsection{Window Shift Operations}
We conduct ablation study to analyze the effect of the window shift operations. The results are reported in Table~\ref{tab:loss}. The results indicate that the window shift operations can build cross-window connections and improve the performance of RFormer. 

\subsection{Clinical Image Analysis and Applications}
\label{ssec:applications}
The ultimate goal of restoring and enhancing fundus images is to serve the real clinical tasks better and improve the accuracy of clinical diagnosis. To validate the effectiveness of our proposed RFormer, we use it as a pre-processing technique for downstream clinical image analysis tasks, including vessel segmentation and optic disc/cup detection. LadderNet~\cite{zhuang2018laddernet} and M-Net~\cite{fu2018joint} are employed as the segmentation baselines.
\subsubsection{User Study}
Since the restored fundus images should meet the requirements of ophthalmologists, we adopt 30 LQ  fundus images for user study. We  use different image restoration methods to enhance these LQ images. Subsequently, we display these results in random order and ask the experienced ophthalmologists to score the quality of the restored images  based on their extensive clinical experience. The score ranges from 0 to 100, larger values are better. The suppression of artifacts and preservation of lesions are taken into account. Finally, we collect responses from five ophthalmologists. The score results for each method are shown in Table.\ref{tab:user}. Our RFormer receives the highest score for best restored results.

\subsubsection{Vessel Segmentation} 
We test LadderNet~\cite{zhuang2018laddernet} pre-trained on DRIVE~\cite{staal2004ridge} dataset for vessel segmentation on our collected RF. Please note that the vessel segmentation maps of real HQ fundus images serve as the references for comparison due to the lack of segmentation labels on our RF. The vessel segmentation results are shown in the third row of Fig.\ref{fenge2}. As can be seen that LadderNet fails in segmenting the blood vessels of clinical LQ fundus images. In contrast, LadderNet extracts obvious vessel structure of the fundus images restored by our RFormer, which is closest to the segmentation results of real HQ fundus images. These results clearly suggest the effectiveness of our proposed method.

\subsubsection{Optic Disc/Cup Detection} 
We also evaluate the effect of our RFormer for the downstream disc/cup detection task. We test M-Net~\cite{fu2018joint} pre-trained on ORIGA~\cite{zhang2010origa} dataset for optic disc/cup detection on our collected RF. Similar to  the vessel segmentation task, the optic disc/cup detection results of real HQ fundus images function as the references due to the lack of segmentation labels. The qualitative comparisons of different fundus image restoration methods are depicted in Fig.~\ref{fenge2}. The fourth line is the optic disc/cup detection map and the fifth line depicts the zoom-in patches of the fourth line. It can be observed that M-Net fails to detect the disc/cup on clinical LQ fundus images. On the contrary, M-Net detects the optic cup and disc more accurately on the fundus images reconstructed by our RFormer. This evidence verifies that our method benefits the optic disc/cup detection task.

\subsection{Fail Cases}
\label{ssec:fail}

Although RFormer achieves good performance, it may not work in some scenes. Fig.~\ref{fig3} shows some fail cases of RFormer on our RF dataset. In the (a) and (e) column, from top to bottom are LQ fundus image, restored fundus image, and HQ fundus image. (b), (c), and (d) are three zoom-in patches of (a). (f), (g), and (h) are three  zoom-in patches of (e). It can be clearly observed from \ref{fig3} (c), (f), and (g) that our RFormer fails to remove the bright spots. As can be seen from \ref{fig3} (b), (d), and (h) that our RFormer fails in enhancing the low-lights regions.  It is difficult for RFormer to learn the feature in areas with insufficient contrast and brightness. We will continue to improve our work according to these fail cases.

\section{Conclusion}
In this paper, we establish the first real clinical fundus image restoration benchmark, Real Fundus, which contains LQ and HQ fundus image pairs to alleviate the data-hungry issue. Our dataset can help better evaluate restoration algorithms in clinical scenes. Based on this dataset, we propose a novel Transformer-based method, RFormer, for clinical fundus image restoration. To the best of our knowledge, it is the first attempt to explore the potential of Transformer in this task. Comprehensive qualitative and quantitative results demonstrate that our RFormer significantly outperforms a series of SOTA methods. Extensive experiments verify that the proposed RFormer serving as a data pre-processing technique can boost the performance of different downstream tasks, such as vessel segmentation and optic disc/cup detection. We hope this work can serve as a baseline for real clinical fundus image restoration and benefit the community of medical  imaging.

\bibliographystyle{IEEEtran}
\bibliography{reference2}

\end{document}